\documentclass[%
 aip,
 amsmath,amssymb,
reprint, 
]{revtex4-1}

\usepackage{graphicx}
\usepackage{dcolumn}
\usepackage{bm}

\usepackage[utf8]{inputenc}
\usepackage[T1]{fontenc}
\usepackage{mathptmx}
\usepackage{etoolbox}
\usepackage{mhchem}
\usepackage{ulem}
\usepackage{todonotes}
\usepackage{siunitx}

\makeatletter
\def\@email#1#2{%
 \endgroup
 \patchcmd{\titleblock@produce}
  {\frontmatter@RRAPformat}
  {\frontmatter@RRAPformat{\produce@RRAP{*#1\href{mailto:#2}{#2}}}\frontmatter@RRAPformat}
  {}{}
}%
\makeatother
\begin{document}

\preprint{AIP/123-QED}

\title[Locally-sealed microfabricated vapor cells filled from an ex situ Cs source]{Locally-sealed microfabricated vapor cells filled from an ex situ Cs source}

\newcommand{\IEMN}{Univ. Lille, CNRS, Centrale Lille, Univ. Polytechnique Hauts-de-France, UMR 8520 - IEMN - Institut d’Electronique de Microélectronique et de Nanotechnologie, Lille, France}
\newcommand{\FEMTO}{Université Marie et Louis Pasteur, CNRS, institut FEMTO-ST, Besançon, France}

\author{L. Péroux}
\affiliation{\IEMN}
\author{A. Dewilde}
\affiliation{\IEMN}
\author{R. Chutani}
\affiliation{\IEMN}
\author{A. Mazzamurro}
\affiliation{\IEMN}
\author{J. Bonhomme}
\affiliation{\IEMN}
\author{A. Mursa}
\affiliation{\FEMTO}
\author{J.-F. Clément}
\affiliation{\IEMN}
\author{A. Talbi}
\affiliation{\IEMN}
\author{P. Pernod}
\affiliation{\IEMN}
\author{N. Passilly}
\affiliation{\FEMTO}
\author{V. Maurice}
\email{vincent.maurice@centralelille.fr}
\affiliation{\IEMN}

\date{\today}

\begin{abstract}
Microfabricated alkali vapor cells are key to enable miniature devices such as atomic clocks and optically pumped magnetometers with reduced size, weight and power.
Yet, more versatile fabrication methods are still needed to further expand their use cases.
Here, we demonstrate a novel approach to collectively fill and seal microfabricated cesium cells using locally-sealed microchannels patterned within one of the glass substrates comprising the cells.
Unlike current methods that rely on wafer-level anodic bonding as the last sealing step, an approach based on local sealing opens the path to features so far limited to traditional glass-blown cells, including the ability to deposit temperature-sensitive antirelaxation coatings, reaching lower background gas pressure without an additional gettering material or filling with diverse atomic or molecular species.
\end{abstract}

\maketitle

The ability to exploit the accurate and stable nature of the atomic structures in practical instruments has opened new pathways in many field applications such as positioning, navigation, timing and sensing.
Benefiting from the ability to build chip-integrated components through micro and nanotechnology processes, chip-scale atomic devices have made significant progress over the past two decades~\cite{kitchingChipscaleAtomicDevices2018}.
In particular, microfabricated alkali vapor cells built at the wafer level have been proposed to achieve small, low-power and, ultimately, accessible devices~\cite{liew_microfabricated_2004}.
Since the first demonstration of an atomic clock exploiting such cells~\cite{knappe_microfabricated_2004-1}, numerous advances have been made.
For instance, integrated optically-pumped magnetometers are now offering an appealing alternative to SQUIDs for magnetoencephalography~\cite{sanderMagnetoencephalographyChipscaleAtomic2012}, optical clocks featuring sub-megahertz linewidths in microfabricated cells have been shown~\cite{newmanArchitecturePhotonicIntegration2019} and quantum memories based on chip-scale cells have recently been proposed~\cite{mottolaOpticalMemoryMicrofabricated2023}.

As opposed to conventional glass-blown cells, whose fabrication remain labor-intensive, prone to variations and difficult to downsize, wafer-based microfabrication processes have allowed making small, repeatable and low-cost cells with the ability to co-integrate features such as coils, heaters and optics~\cite{raghavanFunctionalizedMillimeterscaleVapor2024, chutani_laser_2015, dyerMicromachinedDeepSilicon2022}.
The reported fabrication methods mostly consists in assembling glass and silicon substrates by anodic bonding to form the core of the cell, after having introduced raw alkali metal~\cite{knappe_chip-scale_2005} or a chemical precursor such as Cs/Rb azide or molybdate that can decompose into pure metal after a reaction~\cite{liew_wafer-level_2007, hasegawa_microfabrication_2011-1}.
Over time, improvements have been proposed to scale up and improve the fabrication yields.
For instance, dispensing pure Rb from \ce{BaN6}/\ce{RbCl}, as reported in early chip-scale atomic clocks~\cite{knappe_chip-scale_2005}, has been turned into a wafer-level process with increased throughput~\cite{boppWaferlevelFabricationAlkali2020, liWaferscaleFabricationEvacuated2024}.
Molybdate-based cesium dispensing pills have been demonstrated to fill cells produced on 150-mm-diameter wafers~\cite{vicariniDemonstrationMassproducibleFeature2018} and a paste-like compound has been developed to make the deposition process scalable~\cite{maurice_microfabricated_2017}.


In spite of these advances, the reliance on anodic bonding as the last sealing step in the methods reported above brings limitations that traditional glass-blown cells do not suffer from.
Indeed, anodic bonding releases parasitic gases and prevents thorough outgassing, which limits the purity that can be reached without additional getters~\cite{hasegawa_effects_2013, newmanArchitecturePhotonicIntegration2019, martinezChipscaleAtomicBeam2023}. 
Besides, as anodic bonding involves elevated temperatures (250 to \SI{350}{\celsius}), it precludes the use of temperature-sensitive antirelaxation coatings, which could be beneficial to mitigate the decohering effect of atom-wall collisions~\cite{seltzer_investigation_2010}.
Finally, it also limits the use of gases or molecular species that do not offer practical chemical precursor or that cannot be easily delivered during anodic bonding.

While several workarounds have been studied, no solution as versatile as in conventional cells has been proposed yet.
Low-temperature (\SI{140}{\celsius}) bonding based on indium thermocompression has been investigated, but the reaction between \ce{In} and \ce{Rb} caused the cells to fail prematurely \cite{straessle_microfabricated_2014}.
Cu-Cu thermocompression has also been reported in an attempt to reach a higher purity but the high temperature (\SI{400}{\celsius}) remains unpractical for many use cases and was found to lead to outgassing of residual gases\cite{karlenSealingMEMSAtomic2019}.
We recently proposed a laser-actuated sealing structure based on a glass membrane that deflects to close an underlying channel, thereby locally sealing the inlets of an array of microfabricated cells\cite{mauriceWaferlevelVaporCells2022}.
Despite the given ability to share one alkali source for several cells within the stack, anodic bonding was still used as the last sealing step and dispensers were enclosed within the stack to provide a cesium source.

\begin{figure}
\includegraphics[width=8.5cm]{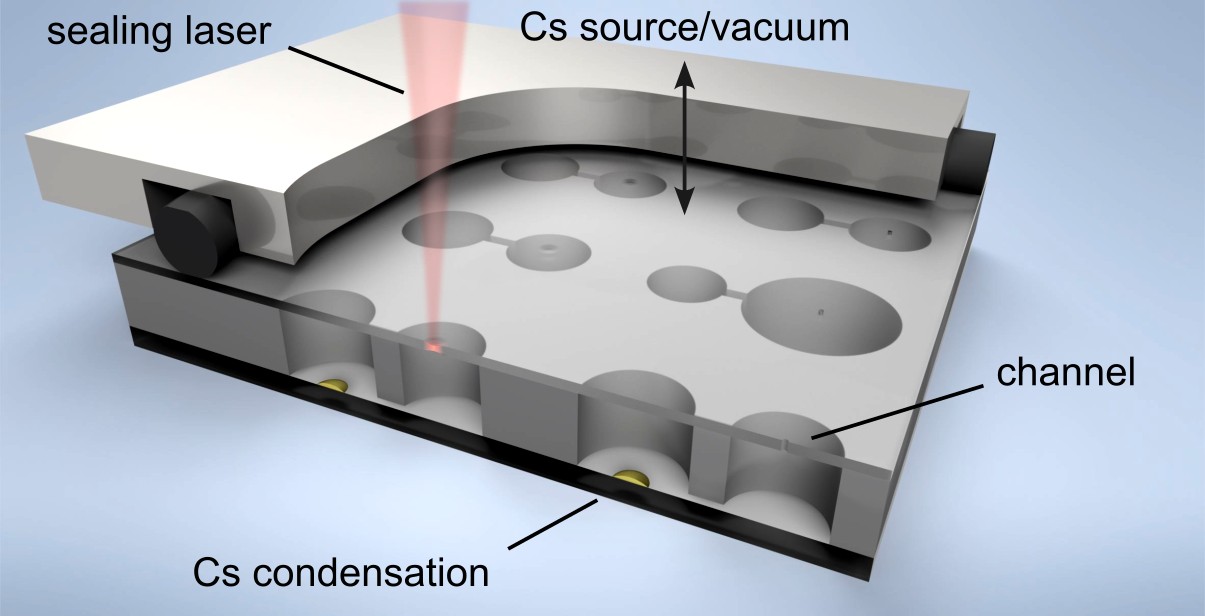}
\caption{\label{fig:principle} Microfabricated cells filled through microchannels from an external cesium source and sealed locally with a \ce{CO2} laser.}
\end{figure}

Here, we demonstrate a significantly simpler approach based on locally-sealed glass channels to fill and seal microfabricated cells from an external source.
Illustrated in Fig.~\ref{fig:principle}, the approach is conceptually similar to the sealing process used in conventional vapor cells and it could offer ways to circumvent some of the limitations that current microfabricated cells are facing.

This approach consists in structuring in the cell's top glass window a channel designed to collapse when locally melted.
Sealing is achieved through surface tensions, which causes the melted glass of the channel's inner surface to flow inward until it eventually closes.
Such channels are patterned at the wafer level and subsequently integrated in the glass/silicon/glass stack constituting the cells.
After migration of cesium into the cell cavity from an external source, the glass channels are sealed by exposure to a focused \ce{CO2} laser beam whose infrared radiation is absorbed by glass.
The cells are fabricated at the wafer level and diced in \SI{19}{\mm}~$\times$~\SI{25}{mm} clusters, each gathering 9 cells.
This form factor allows performing multiple filling/sealing development runs for a single wafer-level fabrication run while still experimenting with parallel filling.
Since the cluster is attached to a vacuum chamber flange during the filling and sealing steps, this cluster size also accommodates smaller size off-the-shelf vacuum components, more practical at an early development stage.
Yet, the process is expected to scale up to allow filling complete 100 mm diameter wafers with custom vacuum components.

\begin{center}
\begin{figure}[h]
\includegraphics{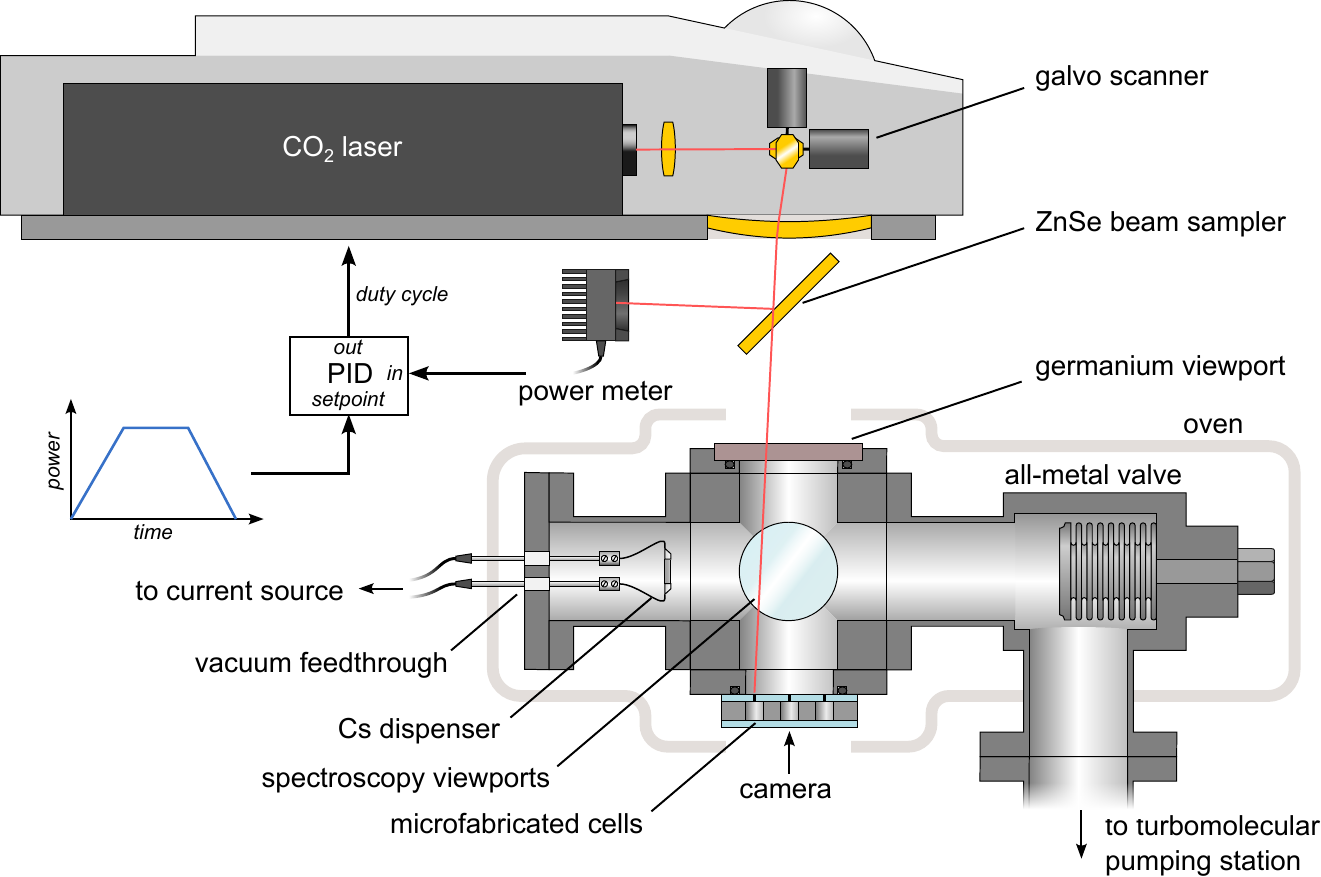}
\caption{\label{fig:schematic} Schematic of the setup used to fill and seal clusters of cells.}
\end{figure}
\end{center}

The setup used to fill and seal the cells is shown in Fig.~\ref{fig:schematic}.
It uses a \ce{CO2} laser head integrating galvanometer scanners (Keyence ML-9110) allowing to steer the laser beam onto each collapsible channel of the cluster.
The laser features a power up to \SI{12}{\watt}, a work area of $\SI{100}{\mm} \times \SI{100}{\mm}$ and a work distance of \SI{190}{\mm}.
Most of the vacuum apparatus comprises components with conflat flanges (CF) joined by Cu gaskets.
The main chamber is a stainless steel cube with 6 CF-40 ports.
The cell cluster is clamped onto a custom-made flange featuring a through-hole facing the glass window with etched channels of the cluster.
The aperture is surrounded by a groove hosting a fluorocarbon rubber (FKM) o-ring and a clamp bolts on top of the cluster to compress the o-ring.
Temporary sealing based on o-rings has previously been employed to study antirelaxation coatings in potassium vapor\cite{seltzer_testing_2008}.
The entire chamber is heated using polyimide flexible heaters clamped onto the flanges.
The microfabricated cell windows facing the outside of the vacuum chamber are chilled with a thermoelectric cooler (TEC) so that temperature gradients favor cesium migration inside the cavities of the cells.
A silicone foam sheet is wrapped around the chamber and the all-metal valve to provide thermal insulation.
The source of alkali metal, here cesium, is a wire-shaped dispenser composed of aluminum, zirconium and cesium salt (Cs/MNF serie from SAES Getters, Lainate, Italy).
One or multiple dispensers can be installed on a flange with electrical feedthroughs.
Two vacuum viewports are installed to provide optical access and perform absorption spectroscopy for assessing the cesium density.
Finally, cesium being highly reactive, notably with \ce{O2} and \ce{H2O}, we aim for the best possible vacuum levels to avoid over-consuming the alkali metal and ensure saturation of the chamber later on.
For this purpose, an all-metal valve connects the chamber to a turbomolecular pumping station and a full range pressure gauge.
\onecolumngrid
\twocolumngrid

\onecolumngrid
\begin{center}
\begin{figure}[h]
\includegraphics{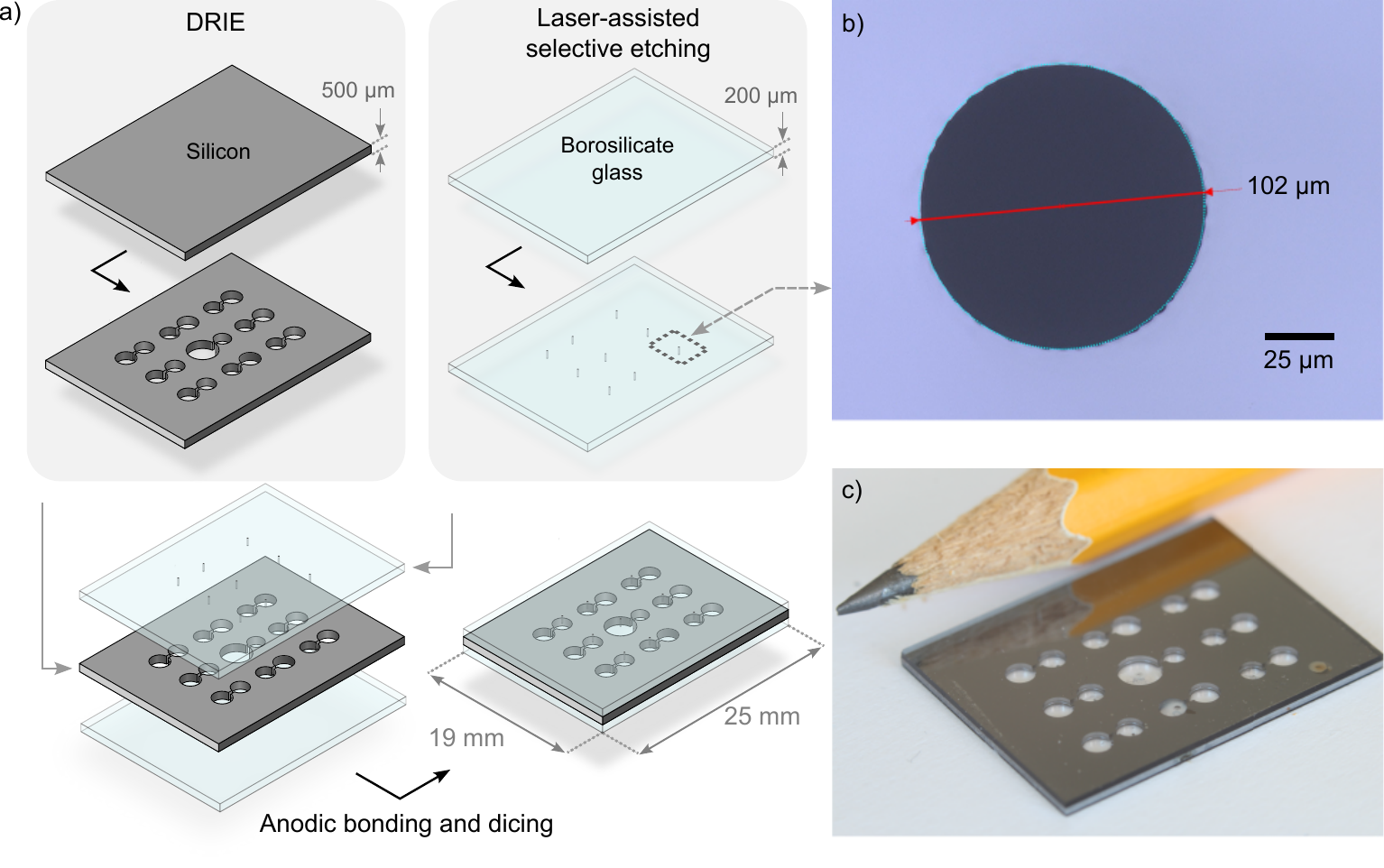}
\caption{\label{fig:flowchart} Microfabrication of cells with collapsible channels. (a) Microfabrication flowchart, each fabrication step in realised on a 4-inch wafer, which is saw-diced into clusters to be compatible with the filling setup. (b) Collapsible channel realised by laser assisted etching with a diameter of \SI{\approx 100}{\um}. (c) Saw-diced cluster of 9 cells. }
\end{figure}
\end{center}
\twocolumngrid

With the cluster of cells placed at the bottom of the cubic vacuum chamber, the \ce{CO2} laser needs to go through the chamber and a vacuum viewport transparent at \SI{10.6}{\um} (wavelength of the laser) must be installed at the top of the chamber.
ZnSe is commonly used in optics for applications operating at this wavelength, however we found it to react and degrade under exposure to warm cesium vapor.
Instead, we turned to a germanium window, which transmits \qty{40}{\percent} of the \SI{10.6}{\um} radiation and up to \qty{95}{\percent} with an antireflective (AR) coating.

Since the laser exhibits large power fluctuations, a servo loop is implemented to stabilize its output power within \qty{1}{\percent} and prevent fast temperature changes, which could generate excessive stress in the glass and hurt repeatability.
For this purpose, a zinc selenide (ZnSe) window angled at \qty{45}{\degree} was added to reflect around \qty{25}{\percent} of the incident power into a power meter.
The cell cluster is imaged by reflecting light from a mirror onto a camera, allowing to precisely aim the laser at the microchannels and monitor the sealing process.

The flowchart describing the microfabrication of the clusters is presented in Fig.~\ref{fig:flowchart}.
The 4-inch wafer design accommodates 11 clusters, each featuring 9 individual cells.
The top substrate of the cells is a \SI{200}{\micro\meter} borosilicate glass wafer, which includes cylindrical channels with a \SI{\sim 100}{\micro\meter} diameter obtained by laser-assisted selective etching, particularly suited to pattern precise geometries (subcontracted from Workshop of Photonics, Vilnius, Lithuania).
The cell cavity is through-etched by deep reactive ion etching (DRIE) in a \SI{500}{\um}~thick silicon wafer.
The silicon wafer is bonded to both the \SI{200}{\um}~thick borosilicate wafer in which the channels have been etched beforehand and a plain \SI{500}{\um}~thick borosilicate wafer by anodic bonding.

After a cluster has been loaded, the chamber is baked under vacuum for a few days at \SI{120}{\celsius}.
The vacuum level, read by a gauge located between the valve and the pumping station, typically reaches \SI{e-7}{mbar}.
After outgassing the dispensers, the all-metal valve is closed, and the dispensers are activated until saturation is observed in the chamber through spectroscopy.
By applying a cold spot with a TEC placed against the outer surface of the cells, a temperature gradient is established, inducing the migration of cesium and the formation of condensation inside the cells (typically within one hour).
After migration, the channels are sequentially locally heated with a \ce{CO2} laser positioned so that the beam width spans \SI{1}{\mm} to seal the cells.
The galvanometer scanners are driven in order to aim precisely at the channel.
The laser power is increased until collapse is observed, usually after a few seconds. Finally the power is slowly ramped down to minimize stress and prevent the formation of cracks.

\begin{figure}
\includegraphics{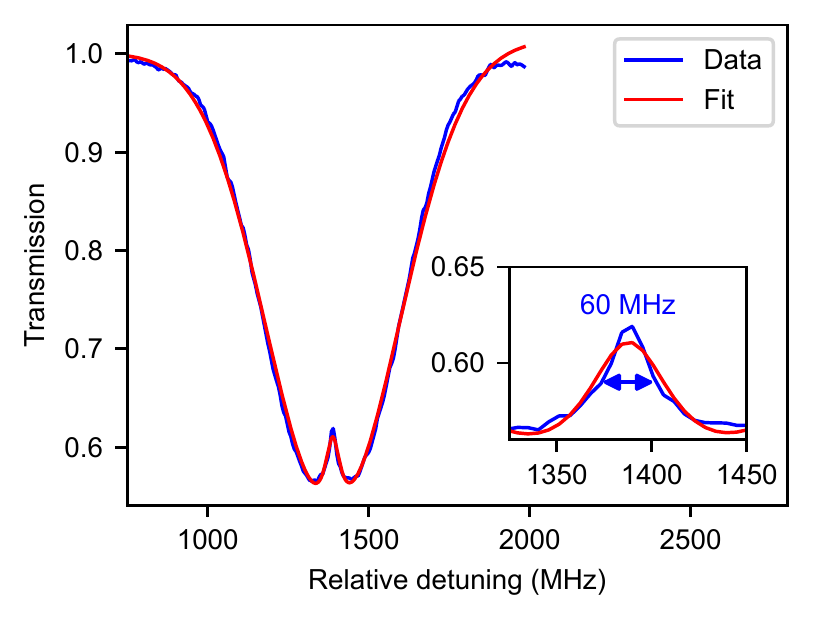}
\caption{\label{fig:abs_sat} Saturated absorption spectrum of the $_{}^{133}\textrm{Cs}$ D1 transition. The measurement was conducted at a temperature of 75°C with a path length of 0.5 mm. The saturated absorption spectra is fitted by a Gaussian line for the absorption peak and a Lorentzian line for the saturated peak, the resulting full width at half maximum (FWHM) is 60~MHz for the Lorentzian.}
\end{figure}

To evaluate the quality of the sealed cells, the residual gas pressure and the stability of the absorption contrast were assessed.
Saturated absorption spectroscopy gives Doppler-free resonances, which helps characterize more finely the influence of the other broadening mechanisms such as residual gas pressure.
The saturated absorption spectra shown in Fig.~\ref{fig:abs_sat} is fitted by a Gaussian for the absorption peak and a Lorentzian for the saturated peak.
The zero-power linewidth $\Gamma_0$ is derived from the linewidth from the Lorentzian fit $\Gamma$ being \SI{60}{\MHz} and the expression $\Gamma = \Gamma_0 \sqrt{1 + I/I_{sat}}$.
Considering the power of the pump beam being \SI[per-mode=symbol]{4.14}{\mW\per\cm\tothe{2}}, we estimate $\Gamma_0$ from which we retrieve the natural linewidth of the $_{}^{133}\textrm{Cs}$ $\mathrm{D_1}$ line $\Gamma_n$, which leads to $\Gamma_0 - \Gamma_n$ equal to \SI{19.89}{\mega\hertz}.
We consider $\mathrm{N_2}$ to be the primary contributor to the residual gas.
Considering a pressure broadening coefficient of \SI{16.55}{\mega\hertz\per Torr} at \SI{75}{\celsius} \cite{pitz_pressure_2009}, we estimate the residual gas pressure to be around \SI{1.2}{Torr}.
This estimation is an upper limit since the zero-power linewidth encompasses both magnetic field broadening and the laser linewidth, which were not taken into account in the calculations.

\begin{figure}
\includegraphics{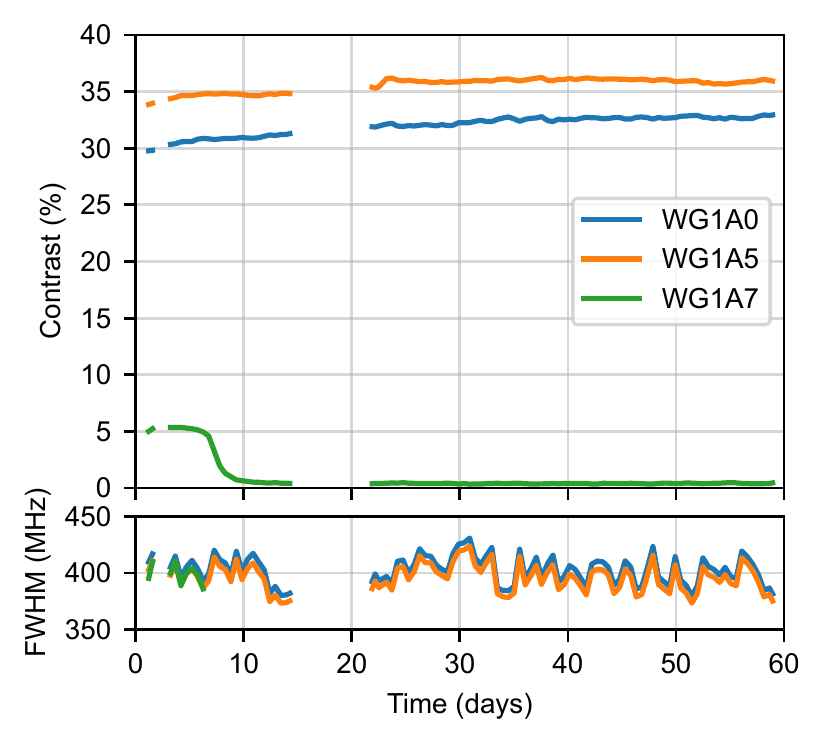}
\caption{\label{fig:long_term} Long-term evolution of the absorption contrast and linewidth. The cluster is kept at a temperature of \SI{70}{\celsius}.}
\end{figure}

The stability of the atomic density was then evaluated to assess the robustness of the sealing over time and ensure that sufficient material was present to maintain a saturated vapor over an extended period.
One cluster of cells has been monitored on an automated linear absorption spectroscopy setup for 60 days at \SI{70}{\celsius} (Fig.~\ref{fig:long_term}).
The presence of cesium was observed in three cells.
One unsaturated cell (WG1A7), whose initial contrast was nearly measured at 5~\%, does not provide absorption anymore.
This cell corresponds to a case where no condensation was migrated.
The two saturated cells (WG1A0, WG1A5), have been showing a stable atmosphere with a contrast ranging between 30 and 40~\% and a constant linewidth at \SI{400}{\mega\hertz}. The contrast difference can be attributed to a variation in temperature.

Compared to the sealing structures based on membranes demonstrated previously \cite{mauriceWaferlevelVaporCells2022}, the method demonstrated here involves a simpler fabrication process.
Unlike membranes, collapsing channels do not rely on a difference of pressure to operate.
They could therefore be used to seal cells with any buffer gas pressure, which is of particular interest to reach high pressures. 
However, the volume of heated glass is larger, which may lead to the consumption of a greater amount of alkali metal after sealing and limit the purity.
The make-seals based on membrane and collapsing channel are complementary and will be both further investigated in the future.


In conclusion, we have presented a novel method for the microfabrication of alkali vapor cells.
It relies on local sealing, making it potentially suitable for the deposition of antirelaxation coatings that would otherwise degrade at the high temperature involved in anodic bonding.
Make-seals based on collapsible glass channels were successfully demonstrated and two cells with saturated Cs vapor were obtained.
Their lifetime is currently being evaluated through long-term linear absorption spectroscopy which has already been stable for 60 days.
Further outgassing could lead to purer cell and make them suitable for applications requiring high vacuum levels.
Developments are ongoing to improve the success rate, the main challenge being the consumption of cesium limiting the duration over which cells can effectively be filled.


\begin{acknowledgments}
We wish to acknowledge the support of Région Hauts-de-France through the Start-AIRR and STaRS programs.
This research has been funded by the French Defence Innovation Agency (Agence de l'Innovation de Défense - AID) and the French Research Agency (Agence Nationale de la Recherche - ANR) within the project ANR-23-ASTR-0020-DA.
This research has also been funded in part by the ANR within the project ANR-23-CE42-0012-01.
This research work has been partially undertaken with the support of IEMN and FEMTO-ST microfabrication facilities (CMNF and MIMENTO), both part of the French Renatech network.
We also acknowledge the support of Centrale Lille's machine shop and technology hub, in particular D. Burgnies, C. Rymek and X. Cimetière.
We thank E. Dordor for the machining work done at FEMTO-ST.
\end{acknowledgments}

\section*{Data Availability Statement}

Data available on request from the authors.



\bibliography{library}

\begin{thebibliography}{24}%
\makeatletter
\providecommand \@ifxundefined [1]{%
 \@ifx{#1\undefined}
}%
\providecommand \@ifnum [1]{%
 \ifnum #1\expandafter \@firstoftwo
 \else \expandafter \@secondoftwo
 \fi
}%
\providecommand \@ifx [1]{%
 \ifx #1\expandafter \@firstoftwo
 \else \expandafter \@secondoftwo
 \fi
}%
\providecommand \natexlab [1]{#1}%
\providecommand \enquote  [1]{``#1''}%
\providecommand \bibnamefont  [1]{#1}%
\providecommand \bibfnamefont [1]{#1}%
\providecommand \citenamefont [1]{#1}%
\providecommand \href@noop [0]{\@secondoftwo}%
\providecommand \href [0]{\begingroup \@sanitize@url \@href}%
\providecommand \@href[1]{\@@startlink{#1}\@@href}%
\providecommand \@@href[1]{\endgroup#1\@@endlink}%
\providecommand \@sanitize@url [0]{\catcode `\\12\catcode `\$12\catcode
  `\&12\catcode `\#12\catcode `\^12\catcode `\_12\catcode `\%12\relax}%
\providecommand \@@startlink[1]{}%
\providecommand \@@endlink[0]{}%
\providecommand \url  [0]{\begingroup\@sanitize@url \@url }%
\providecommand \@url [1]{\endgroup\@href {#1}{\urlprefix }}%
\providecommand \urlprefix  [0]{URL }%
\providecommand \Eprint [0]{\href }%
\providecommand \doibase [0]{http://dx.doi.org/}%
\providecommand \selectlanguage [0]{\@gobble}%
\providecommand \bibinfo  [0]{\@secondoftwo}%
\providecommand \bibfield  [0]{\@secondoftwo}%
\providecommand \translation [1]{[#1]}%
\providecommand \BibitemOpen [0]{}%
\providecommand \bibitemStop [0]{}%
\providecommand \bibitemNoStop [0]{.\EOS\space}%
\providecommand \EOS [0]{\spacefactor3000\relax}%
\providecommand \BibitemShut  [1]{\csname bibitem#1\endcsname}%
\let\auto@bib@innerbib\@empty
\bibitem [{\citenamefont
  {Kitching}(2018)}]{kitchingChipscaleAtomicDevices2018}%
  \BibitemOpen
  \bibfield  {author} {\bibinfo {author} {\bibfnamefont {J.}~\bibnamefont
  {Kitching}},\ }\bibfield  {title} {\enquote {\bibinfo {title} {Chip-scale
  atomic devices},}\ }\href {\doibase 10.1063/1.5026238} {\bibfield  {journal}
  {\bibinfo  {journal} {Applied Physics Reviews}\ }\textbf {\bibinfo {volume}
  {5}},\ \bibinfo {pages} {031302} (\bibinfo {year} {2018})}\BibitemShut
  {NoStop}%
\bibitem [{\citenamefont {Liew}\ \emph {et~al.}(2004)\citenamefont {Liew},
  \citenamefont {Knappe}, \citenamefont {Moreland}, \citenamefont {Robinson},
  \citenamefont {Hollberg},\ and\ \citenamefont
  {Kitching}}]{liew_microfabricated_2004}%
  \BibitemOpen
  \bibfield  {author} {\bibinfo {author} {\bibfnamefont {L.-A.}\ \bibnamefont
  {Liew}}, \bibinfo {author} {\bibfnamefont {S.}~\bibnamefont {Knappe}},
  \bibinfo {author} {\bibfnamefont {J.}~\bibnamefont {Moreland}}, \bibinfo
  {author} {\bibfnamefont {H.~G.}\ \bibnamefont {Robinson}}, \bibinfo {author}
  {\bibfnamefont {L.}~\bibnamefont {Hollberg}}, \ and\ \bibinfo {author}
  {\bibfnamefont {J.}~\bibnamefont {Kitching}},\ }\bibfield  {title} {\enquote
  {\bibinfo {title} {Microfabricated alkali atom vapor cells},}\ }\href
  {\doibase 10.1063/1.1691490} {\bibfield  {journal} {\bibinfo  {journal}
  {Applied Physics Letters}\ }\textbf {\bibinfo {volume} {84}},\ \bibinfo
  {pages} {2694--2696} (\bibinfo {year} {2004})}\BibitemShut {NoStop}%
\bibitem [{\citenamefont {Knappe}\ \emph {et~al.}(2004)\citenamefont {Knappe},
  \citenamefont {Shah}, \citenamefont {Schwindt}, \citenamefont {Hollberg},
  \citenamefont {Kitching}, \citenamefont {Liew},\ and\ \citenamefont
  {Moreland}}]{knappe_microfabricated_2004-1}%
  \BibitemOpen
  \bibfield  {author} {\bibinfo {author} {\bibfnamefont {S.}~\bibnamefont
  {Knappe}}, \bibinfo {author} {\bibfnamefont {V.}~\bibnamefont {Shah}},
  \bibinfo {author} {\bibfnamefont {P.~D.~D.}\ \bibnamefont {Schwindt}},
  \bibinfo {author} {\bibfnamefont {L.}~\bibnamefont {Hollberg}}, \bibinfo
  {author} {\bibfnamefont {J.}~\bibnamefont {Kitching}}, \bibinfo {author}
  {\bibfnamefont {L.-A.}\ \bibnamefont {Liew}}, \ and\ \bibinfo {author}
  {\bibfnamefont {J.}~\bibnamefont {Moreland}},\ }\bibfield  {title} {\enquote
  {\bibinfo {title} {A microfabricated atomic clock},}\ }\href {\doibase
  10.1063/1.1787942} {\bibfield  {journal} {\bibinfo  {journal} {Applied
  Physics Letters}\ }\textbf {\bibinfo {volume} {85}},\ \bibinfo {pages}
  {1460--1462} (\bibinfo {year} {2004})}\BibitemShut {NoStop}%
\bibitem [{\citenamefont {Sander}\ \emph {et~al.}(2012)\citenamefont {Sander},
  \citenamefont {Preusser}, \citenamefont {Mhaskar}, \citenamefont {Kitching},
  \citenamefont {Trahms},\ and\ \citenamefont
  {Knappe}}]{sanderMagnetoencephalographyChipscaleAtomic2012}%
  \BibitemOpen
  \bibfield  {author} {\bibinfo {author} {\bibfnamefont {T.~H.}\ \bibnamefont
  {Sander}}, \bibinfo {author} {\bibfnamefont {J.}~\bibnamefont {Preusser}},
  \bibinfo {author} {\bibfnamefont {R.~R.}\ \bibnamefont {Mhaskar}}, \bibinfo
  {author} {\bibfnamefont {J.}~\bibnamefont {Kitching}}, \bibinfo {author}
  {\bibfnamefont {L.}~\bibnamefont {Trahms}}, \ and\ \bibinfo {author}
  {\bibfnamefont {S.}~\bibnamefont {Knappe}},\ }\bibfield  {title} {\enquote
  {\bibinfo {title} {Magnetoencephalography with a chip-scale atomic
  magnetometer},}\ }\href {\doibase 10.1364/BOE.3.000981} {\bibfield  {journal}
  {\bibinfo  {journal} {Optics Express}\ }\textbf {\bibinfo {volume} {3}},\
  \bibinfo {pages} {27167--27172} (\bibinfo {year} {2012})}\BibitemShut
  {NoStop}%
\bibitem [{\citenamefont {Newman}\ \emph {et~al.}(2019)\citenamefont {Newman},
  \citenamefont {Maurice}, \citenamefont {Drake}, \citenamefont {Stone},
  \citenamefont {Briles}, \citenamefont {Spencer}, \citenamefont {Fredrick},
  \citenamefont {Li}, \citenamefont {Westly}, \citenamefont {Ilic},
  \citenamefont {Shen}, \citenamefont {Suh}, \citenamefont {Yang},
  \citenamefont {Johnson}, \citenamefont {Johnson}, \citenamefont {Hollberg},
  \citenamefont {Vahala}, \citenamefont {Srinivasan}, \citenamefont {Diddams},
  \citenamefont {Kitching}, \citenamefont {Papp},\ and\ \citenamefont
  {Hummon}}]{newmanArchitecturePhotonicIntegration2019}%
  \BibitemOpen
  \bibfield  {author} {\bibinfo {author} {\bibfnamefont {Z.~L.}\ \bibnamefont
  {Newman}}, \bibinfo {author} {\bibfnamefont {V.}~\bibnamefont {Maurice}},
  \bibinfo {author} {\bibfnamefont {T.}~\bibnamefont {Drake}}, \bibinfo
  {author} {\bibfnamefont {J.~R.}\ \bibnamefont {Stone}}, \bibinfo {author}
  {\bibfnamefont {T.~C.}\ \bibnamefont {Briles}}, \bibinfo {author}
  {\bibfnamefont {D.~T.}\ \bibnamefont {Spencer}}, \bibinfo {author}
  {\bibfnamefont {C.}~\bibnamefont {Fredrick}}, \bibinfo {author}
  {\bibfnamefont {Q.}~\bibnamefont {Li}}, \bibinfo {author} {\bibfnamefont
  {D.}~\bibnamefont {Westly}}, \bibinfo {author} {\bibfnamefont {B.~R.}\
  \bibnamefont {Ilic}}, \bibinfo {author} {\bibfnamefont {B.}~\bibnamefont
  {Shen}}, \bibinfo {author} {\bibfnamefont {M.-G.}\ \bibnamefont {Suh}},
  \bibinfo {author} {\bibfnamefont {K.~Y.}\ \bibnamefont {Yang}}, \bibinfo
  {author} {\bibfnamefont {C.}~\bibnamefont {Johnson}}, \bibinfo {author}
  {\bibfnamefont {D.~M.~S.}\ \bibnamefont {Johnson}}, \bibinfo {author}
  {\bibfnamefont {L.}~\bibnamefont {Hollberg}}, \bibinfo {author}
  {\bibfnamefont {K.~J.}\ \bibnamefont {Vahala}}, \bibinfo {author}
  {\bibfnamefont {K.}~\bibnamefont {Srinivasan}}, \bibinfo {author}
  {\bibfnamefont {S.~A.}\ \bibnamefont {Diddams}}, \bibinfo {author}
  {\bibfnamefont {J.}~\bibnamefont {Kitching}}, \bibinfo {author}
  {\bibfnamefont {S.~B.}\ \bibnamefont {Papp}}, \ and\ \bibinfo {author}
  {\bibfnamefont {M.~T.}\ \bibnamefont {Hummon}},\ }\bibfield  {title}
  {\enquote {\bibinfo {title} {Architecture for the photonic integration of an
  optical atomic clock},}\ }\href {\doibase 10.1364/OPTICA.6.000680} {\bibfield
   {journal} {\bibinfo  {journal} {Optica}\ }\textbf {\bibinfo {volume} {6}},\
  \bibinfo {pages} {680--685} (\bibinfo {year} {2019})}\BibitemShut {NoStop}%
\bibitem [{\citenamefont {Mottola}, \citenamefont {Buser},\ and\ \citenamefont
  {Treutlein}(2023)}]{mottolaOpticalMemoryMicrofabricated2023}%
  \BibitemOpen
  \bibfield  {author} {\bibinfo {author} {\bibfnamefont {R.}~\bibnamefont
  {Mottola}}, \bibinfo {author} {\bibfnamefont {G.}~\bibnamefont {Buser}}, \
  and\ \bibinfo {author} {\bibfnamefont {P.}~\bibnamefont {Treutlein}},\
  }\bibfield  {title} {\enquote {\bibinfo {title} {Optical {{Memory}} in a
  {{Microfabricated Rubidium Vapor Cell}}},}\ }\href {\doibase
  10.1103/PhysRevLett.131.260801} {\bibfield  {journal} {\bibinfo  {journal}
  {Physical Review Letters}\ }\textbf {\bibinfo {volume} {131}},\ \bibinfo
  {pages} {260801} (\bibinfo {year} {2023})}\BibitemShut {NoStop}%
\bibitem [{\citenamefont {Raghavan}\ \emph {et~al.}(2024)\citenamefont
  {Raghavan}, \citenamefont {Tayler}, \citenamefont {Mouloudakis},
  \citenamefont {Rae}, \citenamefont {L{\"a}hteenm{\"a}ki}, \citenamefont
  {Zetter}, \citenamefont {Laine}, \citenamefont {Haesler}, \citenamefont
  {Balet}, \citenamefont {Overstolz}, \citenamefont {Karlen},\ and\
  \citenamefont {Mitchell}}]{raghavanFunctionalizedMillimeterscaleVapor2024}%
  \BibitemOpen
  \bibfield  {author} {\bibinfo {author} {\bibfnamefont {H.}~\bibnamefont
  {Raghavan}}, \bibinfo {author} {\bibfnamefont {M.~C.}\ \bibnamefont
  {Tayler}}, \bibinfo {author} {\bibfnamefont {K.}~\bibnamefont {Mouloudakis}},
  \bibinfo {author} {\bibfnamefont {R.}~\bibnamefont {Rae}}, \bibinfo {author}
  {\bibfnamefont {S.}~\bibnamefont {L{\"a}hteenm{\"a}ki}}, \bibinfo {author}
  {\bibfnamefont {R.}~\bibnamefont {Zetter}}, \bibinfo {author} {\bibfnamefont
  {P.}~\bibnamefont {Laine}}, \bibinfo {author} {\bibfnamefont
  {J.}~\bibnamefont {Haesler}}, \bibinfo {author} {\bibfnamefont
  {L.}~\bibnamefont {Balet}}, \bibinfo {author} {\bibfnamefont
  {T.}~\bibnamefont {Overstolz}}, \bibinfo {author} {\bibfnamefont
  {S.}~\bibnamefont {Karlen}}, \ and\ \bibinfo {author} {\bibfnamefont {M.~W.}\
  \bibnamefont {Mitchell}},\ }\bibfield  {title} {\enquote {\bibinfo {title}
  {Functionalized millimeter-scale vapor cells for alkali-metal spectroscopy
  and magnetometry},}\ }\href {\doibase 10.1103/PhysRevApplied.22.044011}
  {\bibfield  {journal} {\bibinfo  {journal} {Physical Review Applied}\
  }\textbf {\bibinfo {volume} {22}},\ \bibinfo {pages} {044011} (\bibinfo
  {year} {2024})}\BibitemShut {NoStop}%
\bibitem [{\citenamefont {Chutani}\ \emph {et~al.}(2015)\citenamefont
  {Chutani}, \citenamefont {Maurice}, \citenamefont {Passilly}, \citenamefont
  {Gorecki}, \citenamefont {Boudot}, \citenamefont {Abdel~Hafiz}, \citenamefont
  {Abb{\'e}}, \citenamefont {Galliou}, \citenamefont {Rauch},\ and\
  \citenamefont {{de Clercq}}}]{chutani_laser_2015}%
  \BibitemOpen
  \bibfield  {author} {\bibinfo {author} {\bibfnamefont {R.~K.}\ \bibnamefont
  {Chutani}}, \bibinfo {author} {\bibfnamefont {V.}~\bibnamefont {Maurice}},
  \bibinfo {author} {\bibfnamefont {N.}~\bibnamefont {Passilly}}, \bibinfo
  {author} {\bibfnamefont {C.}~\bibnamefont {Gorecki}}, \bibinfo {author}
  {\bibfnamefont {R.}~\bibnamefont {Boudot}}, \bibinfo {author} {\bibfnamefont
  {M.}~\bibnamefont {Abdel~Hafiz}}, \bibinfo {author} {\bibfnamefont
  {P.}~\bibnamefont {Abb{\'e}}}, \bibinfo {author} {\bibfnamefont
  {S.}~\bibnamefont {Galliou}}, \bibinfo {author} {\bibfnamefont {J.-Y.}\
  \bibnamefont {Rauch}}, \ and\ \bibinfo {author} {\bibfnamefont
  {E.}~\bibnamefont {{de Clercq}}},\ }\bibfield  {title} {\enquote {\bibinfo
  {title} {Laser light routing in an elongated micromachined vapor cell with
  diffraction gratings for atomic clock applications},}\ }\href {\doibase
  10.1038/srep14001} {\bibfield  {journal} {\bibinfo  {journal} {Scientific
  Reports}\ }\textbf {\bibinfo {volume} {5}},\ \bibinfo {pages} {14001}
  (\bibinfo {year} {2015})}\BibitemShut {NoStop}%
\bibitem [{\citenamefont {Dyer}\ \emph {et~al.}(2022)\citenamefont {Dyer},
  \citenamefont {Griffin}, \citenamefont {Arnold}, \citenamefont {Mirando},
  \citenamefont {Burt}, \citenamefont {Riis},\ and\ \citenamefont
  {McGilligan}}]{dyerMicromachinedDeepSilicon2022}%
  \BibitemOpen
  \bibfield  {author} {\bibinfo {author} {\bibfnamefont {S.}~\bibnamefont
  {Dyer}}, \bibinfo {author} {\bibfnamefont {P.~F.}\ \bibnamefont {Griffin}},
  \bibinfo {author} {\bibfnamefont {A.~S.}\ \bibnamefont {Arnold}}, \bibinfo
  {author} {\bibfnamefont {F.}~\bibnamefont {Mirando}}, \bibinfo {author}
  {\bibfnamefont {D.~P.}\ \bibnamefont {Burt}}, \bibinfo {author}
  {\bibfnamefont {E.}~\bibnamefont {Riis}}, \ and\ \bibinfo {author}
  {\bibfnamefont {J.~P.}\ \bibnamefont {McGilligan}},\ }\bibfield  {title}
  {\enquote {\bibinfo {title} {Micro-machined deep silicon atomic vapor
  cells},}\ }\href {\doibase 10.1063/5.0114762} {\bibfield  {journal} {\bibinfo
   {journal} {Journal of Applied Physics}\ }\textbf {\bibinfo {volume} {132}},\
  \bibinfo {pages} {134401} (\bibinfo {year} {2022})}\BibitemShut {NoStop}%
\bibitem [{\citenamefont {Knappe}\ \emph {et~al.}(2005)\citenamefont {Knappe},
  \citenamefont {Schwindt}, \citenamefont {Shah}, \citenamefont {Hollberg},
  \citenamefont {Kitching}, \citenamefont {Liew},\ and\ \citenamefont
  {Moreland}}]{knappe_chip-scale_2005}%
  \BibitemOpen
  \bibfield  {author} {\bibinfo {author} {\bibfnamefont {S.}~\bibnamefont
  {Knappe}}, \bibinfo {author} {\bibfnamefont {P.~D.~D.}\ \bibnamefont
  {Schwindt}}, \bibinfo {author} {\bibfnamefont {V.}~\bibnamefont {Shah}},
  \bibinfo {author} {\bibfnamefont {L.}~\bibnamefont {Hollberg}}, \bibinfo
  {author} {\bibfnamefont {J.}~\bibnamefont {Kitching}}, \bibinfo {author}
  {\bibfnamefont {L.-A.}\ \bibnamefont {Liew}}, \ and\ \bibinfo {author}
  {\bibfnamefont {J.}~\bibnamefont {Moreland}},\ }\bibfield  {title} {\enquote
  {\bibinfo {title} {A chip-scale atomic clock based on $^{87}${Rb} with
  improved frequency stability},}\ }\href {\doibase 10.1364/OPEX.13.001249}
  {\bibfield  {journal} {\bibinfo  {journal} {Optics Express}\ }\textbf
  {\bibinfo {volume} {13}},\ \bibinfo {pages} {1249--1253} (\bibinfo {year}
  {2005})},\ \Eprint {http://arxiv.org/abs/19494996} {19494996} \BibitemShut
  {NoStop}%
\bibitem [{\citenamefont {Liew}, \citenamefont {Moreland},\ and\ \citenamefont
  {Gerginov}(2007)}]{liew_wafer-level_2007}%
  \BibitemOpen
  \bibfield  {author} {\bibinfo {author} {\bibfnamefont {L.-A.}\ \bibnamefont
  {Liew}}, \bibinfo {author} {\bibfnamefont {J.}~\bibnamefont {Moreland}}, \
  and\ \bibinfo {author} {\bibfnamefont {V.}~\bibnamefont {Gerginov}},\
  }\bibfield  {title} {\enquote {\bibinfo {title} {Wafer-level filling of
  microfabricated atomic vapor cells based on thin-film deposition and
  photolysis of cesium azide},}\ }\href {\doibase 10.1063/1.2712501} {\bibfield
   {journal} {\bibinfo  {journal} {Applied Physics Letters}\ }\textbf {\bibinfo
  {volume} {90}},\ \bibinfo {pages} {114106} (\bibinfo {year}
  {2007})}\BibitemShut {NoStop}%
\bibitem [{\citenamefont {Hasegawa}\ \emph {et~al.}(2011)\citenamefont
  {Hasegawa}, \citenamefont {Chutani}, \citenamefont {Gorecki}, \citenamefont
  {Boudot}, \citenamefont {Dziuban}, \citenamefont {Giordano}, \citenamefont
  {Clatot},\ and\ \citenamefont {Mauri}}]{hasegawa_microfabrication_2011-1}%
  \BibitemOpen
  \bibfield  {author} {\bibinfo {author} {\bibfnamefont {M.}~\bibnamefont
  {Hasegawa}}, \bibinfo {author} {\bibfnamefont {R.~K.}\ \bibnamefont
  {Chutani}}, \bibinfo {author} {\bibfnamefont {C.}~\bibnamefont {Gorecki}},
  \bibinfo {author} {\bibfnamefont {R.}~\bibnamefont {Boudot}}, \bibinfo
  {author} {\bibfnamefont {P.}~\bibnamefont {Dziuban}}, \bibinfo {author}
  {\bibfnamefont {V.}~\bibnamefont {Giordano}}, \bibinfo {author}
  {\bibfnamefont {S.}~\bibnamefont {Clatot}}, \ and\ \bibinfo {author}
  {\bibfnamefont {L.}~\bibnamefont {Mauri}},\ }\bibfield  {title} {\enquote
  {\bibinfo {title} {Microfabrication of cesium vapor cells with buffer gas for
  {{MEMS}} atomic clocks},}\ }\href {\doibase 10.1016/j.sna.2011.02.039}
  {\bibfield  {journal} {\bibinfo  {journal} {Sensors and Actuators A:
  Physical}\ }\textbf {\bibinfo {volume} {167}},\ \bibinfo {pages} {594--601}
  (\bibinfo {year} {2011})}\BibitemShut {NoStop}%
\bibitem [{\citenamefont {Bopp}, \citenamefont {Maurice},\ and\ \citenamefont
  {Kitching}(2020)}]{boppWaferlevelFabricationAlkali2020}%
  \BibitemOpen
  \bibfield  {author} {\bibinfo {author} {\bibfnamefont {D.~G.}\ \bibnamefont
  {Bopp}}, \bibinfo {author} {\bibfnamefont {V.~M.}\ \bibnamefont {Maurice}}, \
  and\ \bibinfo {author} {\bibfnamefont {J.~E.}\ \bibnamefont {Kitching}},\
  }\bibfield  {title} {\enquote {\bibinfo {title} {Wafer-level fabrication of
  alkali vapor cells using in-situ atomic deposition},}\ }\href {\doibase
  10.1088/2515-7647/abcbe5} {\bibfield  {journal} {\bibinfo  {journal} {Journal
  of Physics: Photonics}\ }\textbf {\bibinfo {volume} {3}},\ \bibinfo {pages}
  {015002} (\bibinfo {year} {2020})}\BibitemShut {NoStop}%
\bibitem [{\citenamefont {Li}\ \emph {et~al.}(2024)\citenamefont {Li},
  \citenamefont {Sohn}, \citenamefont {Hummon}, \citenamefont {Schima},\ and\
  \citenamefont {Kitching}}]{liWaferscaleFabricationEvacuated2024}%
  \BibitemOpen
  \bibfield  {author} {\bibinfo {author} {\bibfnamefont {Y.}~\bibnamefont
  {Li}}, \bibinfo {author} {\bibfnamefont {D.~B.}\ \bibnamefont {Sohn}},
  \bibinfo {author} {\bibfnamefont {M.~T.}\ \bibnamefont {Hummon}}, \bibinfo
  {author} {\bibfnamefont {S.}~\bibnamefont {Schima}}, \ and\ \bibinfo {author}
  {\bibfnamefont {J.}~\bibnamefont {Kitching}},\ }\bibfield  {title} {\enquote
  {\bibinfo {title} {Wafer-scale fabrication of evacuated alkali vapor
  cells},}\ }\href {\doibase 10.1364/OL.527351} {\bibfield  {journal} {\bibinfo
   {journal} {Optics Letters}\ }\textbf {\bibinfo {volume} {49}},\ \bibinfo
  {pages} {4963--4966} (\bibinfo {year} {2024})}\BibitemShut {NoStop}%
\bibitem [{\citenamefont {Vicarini}\ \emph {et~al.}(2018)\citenamefont
  {Vicarini}, \citenamefont {Maurice}, \citenamefont {Abdel~Hafiz},
  \citenamefont {Rutkowski}, \citenamefont {Gorecki}, \citenamefont {Passilly},
  \citenamefont {Ribetto}, \citenamefont {Gaff}, \citenamefont {Volant},
  \citenamefont {Galliou},\ and\ \citenamefont
  {Boudot}}]{vicariniDemonstrationMassproducibleFeature2018}%
  \BibitemOpen
  \bibfield  {author} {\bibinfo {author} {\bibfnamefont {R.}~\bibnamefont
  {Vicarini}}, \bibinfo {author} {\bibfnamefont {V.}~\bibnamefont {Maurice}},
  \bibinfo {author} {\bibfnamefont {M.}~\bibnamefont {Abdel~Hafiz}}, \bibinfo
  {author} {\bibfnamefont {J.}~\bibnamefont {Rutkowski}}, \bibinfo {author}
  {\bibfnamefont {C.}~\bibnamefont {Gorecki}}, \bibinfo {author} {\bibfnamefont
  {N.}~\bibnamefont {Passilly}}, \bibinfo {author} {\bibfnamefont
  {L.}~\bibnamefont {Ribetto}}, \bibinfo {author} {\bibfnamefont
  {V.}~\bibnamefont {Gaff}}, \bibinfo {author} {\bibfnamefont {V.}~\bibnamefont
  {Volant}}, \bibinfo {author} {\bibfnamefont {S.}~\bibnamefont {Galliou}}, \
  and\ \bibinfo {author} {\bibfnamefont {R.}~\bibnamefont {Boudot}},\
  }\bibfield  {title} {\enquote {\bibinfo {title} {Demonstration of the
  mass-producible feature of a {{Cs}} vapor microcell technology for miniature
  atomic clocks},}\ }\href {\doibase 10.1016/j.sna.2018.07.032} {\bibfield
  {journal} {\bibinfo  {journal} {Sensors and Actuators A: Physical}\ }\textbf
  {\bibinfo {volume} {280}},\ \bibinfo {pages} {99--106} (\bibinfo {year}
  {2018})}\BibitemShut {NoStop}%
\bibitem [{\citenamefont {Maurice}\ \emph {et~al.}(2017)\citenamefont
  {Maurice}, \citenamefont {Rutkowski}, \citenamefont {Kroemer}, \citenamefont
  {Bargiel}, \citenamefont {Passilly}, \citenamefont {Boudot}, \citenamefont
  {Gorecki}, \citenamefont {Mauri},\ and\ \citenamefont
  {Moraja}}]{maurice_microfabricated_2017}%
  \BibitemOpen
  \bibfield  {author} {\bibinfo {author} {\bibfnamefont {V.}~\bibnamefont
  {Maurice}}, \bibinfo {author} {\bibfnamefont {J.}~\bibnamefont {Rutkowski}},
  \bibinfo {author} {\bibfnamefont {E.}~\bibnamefont {Kroemer}}, \bibinfo
  {author} {\bibfnamefont {S.}~\bibnamefont {Bargiel}}, \bibinfo {author}
  {\bibfnamefont {N.}~\bibnamefont {Passilly}}, \bibinfo {author}
  {\bibfnamefont {R.}~\bibnamefont {Boudot}}, \bibinfo {author} {\bibfnamefont
  {C.}~\bibnamefont {Gorecki}}, \bibinfo {author} {\bibfnamefont
  {L.}~\bibnamefont {Mauri}}, \ and\ \bibinfo {author} {\bibfnamefont
  {M.}~\bibnamefont {Moraja}},\ }\bibfield  {title} {\enquote {\bibinfo {title}
  {Microfabricated vapor cells filled with a cesium dispensing paste for
  miniature atomic clocks},}\ }\href {\doibase 10.1063/1.4981772} {\bibfield
  {journal} {\bibinfo  {journal} {Applied Physics Letters}\ }\textbf {\bibinfo
  {volume} {110}},\ \bibinfo {pages} {164103} (\bibinfo {year}
  {2017})}\BibitemShut {NoStop}%
\bibitem [{\citenamefont {Hasegawa}\ \emph {et~al.}(2013)\citenamefont
  {Hasegawa}, \citenamefont {Chutani}, \citenamefont {Boudot}, \citenamefont
  {Mauri}, \citenamefont {Gorecki}, \citenamefont {Liu},\ and\ \citenamefont
  {Passilly}}]{hasegawa_effects_2013}%
  \BibitemOpen
  \bibfield  {author} {\bibinfo {author} {\bibfnamefont {M.}~\bibnamefont
  {Hasegawa}}, \bibinfo {author} {\bibfnamefont {R.~K.}\ \bibnamefont
  {Chutani}}, \bibinfo {author} {\bibfnamefont {R.}~\bibnamefont {Boudot}},
  \bibinfo {author} {\bibfnamefont {L.}~\bibnamefont {Mauri}}, \bibinfo
  {author} {\bibfnamefont {C.}~\bibnamefont {Gorecki}}, \bibinfo {author}
  {\bibfnamefont {X.}~\bibnamefont {Liu}}, \ and\ \bibinfo {author}
  {\bibfnamefont {N.}~\bibnamefont {Passilly}},\ }\bibfield  {title} {\enquote
  {\bibinfo {title} {Effects of getters on hermetically sealed micromachined
  cesium--neon cells for atomic clocks},}\ }\href {\doibase
  10.1088/0960-1317/23/5/055022} {\bibfield  {journal} {\bibinfo  {journal}
  {Journal of Micromechanics and Microengineering}\ }\textbf {\bibinfo {volume}
  {23}},\ \bibinfo {pages} {055022} (\bibinfo {year} {2013})}\BibitemShut
  {NoStop}%
\bibitem [{\citenamefont {Martinez}\ \emph {et~al.}(2023)\citenamefont
  {Martinez}, \citenamefont {Li}, \citenamefont {Staron}, \citenamefont
  {Kitching}, \citenamefont {Raman},\ and\ \citenamefont
  {McGehee}}]{martinezChipscaleAtomicBeam2023}%
  \BibitemOpen
  \bibfield  {author} {\bibinfo {author} {\bibfnamefont {G.~D.}\ \bibnamefont
  {Martinez}}, \bibinfo {author} {\bibfnamefont {C.}~\bibnamefont {Li}},
  \bibinfo {author} {\bibfnamefont {A.}~\bibnamefont {Staron}}, \bibinfo
  {author} {\bibfnamefont {J.}~\bibnamefont {Kitching}}, \bibinfo {author}
  {\bibfnamefont {C.}~\bibnamefont {Raman}}, \ and\ \bibinfo {author}
  {\bibfnamefont {W.~R.}\ \bibnamefont {McGehee}},\ }\bibfield  {title}
  {\enquote {\bibinfo {title} {A chip-scale atomic beam clock},}\ }\href
  {\doibase 10.1038/s41467-023-39166-1} {\bibfield  {journal} {\bibinfo
  {journal} {Nature Communications}\ }\textbf {\bibinfo {volume} {14}},\
  \bibinfo {pages} {3501} (\bibinfo {year} {2023})}\BibitemShut {NoStop}%
\bibitem [{\citenamefont {Seltzer}\ \emph {et~al.}(2010)\citenamefont
  {Seltzer}, \citenamefont {Michalak}, \citenamefont {Donaldson}, \citenamefont
  {Balabas}, \citenamefont {Barber}, \citenamefont {Bernasek}, \citenamefont
  {Bouchiat}, \citenamefont {Hexemer}, \citenamefont {Hibberd}, \citenamefont
  {Kimball}, \citenamefont {Jaye}, \citenamefont {Karaulanov}, \citenamefont
  {Narducci}, \citenamefont {Rangwala}, \citenamefont {Robinson}, \citenamefont
  {Shmakov}, \citenamefont {Voronov}, \citenamefont {Yashchuk}, \citenamefont
  {Pines},\ and\ \citenamefont {Budker}}]{seltzer_investigation_2010}%
  \BibitemOpen
  \bibfield  {author} {\bibinfo {author} {\bibfnamefont {S.~J.}\ \bibnamefont
  {Seltzer}}, \bibinfo {author} {\bibfnamefont {D.~J.}\ \bibnamefont
  {Michalak}}, \bibinfo {author} {\bibfnamefont {M.~H.}\ \bibnamefont
  {Donaldson}}, \bibinfo {author} {\bibfnamefont {M.~V.}\ \bibnamefont
  {Balabas}}, \bibinfo {author} {\bibfnamefont {S.~K.}\ \bibnamefont {Barber}},
  \bibinfo {author} {\bibfnamefont {S.~L.}\ \bibnamefont {Bernasek}}, \bibinfo
  {author} {\bibfnamefont {M.-A.}\ \bibnamefont {Bouchiat}}, \bibinfo {author}
  {\bibfnamefont {A.}~\bibnamefont {Hexemer}}, \bibinfo {author} {\bibfnamefont
  {{\relax AM}.}~\bibnamefont {Hibberd}}, \bibinfo {author} {\bibfnamefont
  {D.~F.~J.}\ \bibnamefont {Kimball}}, \bibinfo {author} {\bibfnamefont
  {C.}~\bibnamefont {Jaye}}, \bibinfo {author} {\bibfnamefont {T.}~\bibnamefont
  {Karaulanov}}, \bibinfo {author} {\bibfnamefont {F.~A.}\ \bibnamefont
  {Narducci}}, \bibinfo {author} {\bibfnamefont {S.~A.}\ \bibnamefont
  {Rangwala}}, \bibinfo {author} {\bibfnamefont {H.~G.}\ \bibnamefont
  {Robinson}}, \bibinfo {author} {\bibfnamefont {A.~K.}\ \bibnamefont
  {Shmakov}}, \bibinfo {author} {\bibfnamefont {D.~L.}\ \bibnamefont
  {Voronov}}, \bibinfo {author} {\bibfnamefont {V.~V.}\ \bibnamefont
  {Yashchuk}}, \bibinfo {author} {\bibfnamefont {A.}~\bibnamefont {Pines}}, \
  and\ \bibinfo {author} {\bibfnamefont {D.}~\bibnamefont {Budker}},\
  }\bibfield  {title} {\enquote {\bibinfo {title} {Investigation of
  antirelaxation coatings for alkali-metal vapor cells using surface science
  techniques.}}\ }\href {\doibase 10.1063/1.3489922} {\bibfield  {journal}
  {\bibinfo  {journal} {The Journal of Chemical Physics}\ }\textbf {\bibinfo
  {volume} {133}},\ \bibinfo {pages} {144703} (\bibinfo {year} {2010})},\
  \Eprint {http://arxiv.org/abs/20950026} {20950026} \BibitemShut {NoStop}%
\bibitem [{\citenamefont {Straessle}\ \emph {et~al.}(2014)\citenamefont
  {Straessle}, \citenamefont {Pellaton}, \citenamefont {Affolderbach},
  \citenamefont {P{\'e}tremand}, \citenamefont {Briand}, \citenamefont
  {Mileti},\ and\ \citenamefont {{de Rooij}}}]{straessle_microfabricated_2014}%
  \BibitemOpen
  \bibfield  {author} {\bibinfo {author} {\bibfnamefont {R.}~\bibnamefont
  {Straessle}}, \bibinfo {author} {\bibfnamefont {M.}~\bibnamefont {Pellaton}},
  \bibinfo {author} {\bibfnamefont {C.}~\bibnamefont {Affolderbach}}, \bibinfo
  {author} {\bibfnamefont {Y.}~\bibnamefont {P{\'e}tremand}}, \bibinfo {author}
  {\bibfnamefont {D.}~\bibnamefont {Briand}}, \bibinfo {author} {\bibfnamefont
  {G.}~\bibnamefont {Mileti}}, \ and\ \bibinfo {author} {\bibfnamefont {N.~F.}\
  \bibnamefont {{de Rooij}}},\ }\bibfield  {title} {\enquote {\bibinfo {title}
  {Microfabricated alkali vapor cell with anti-relaxation wall coating},}\
  }\href {\doibase 10.1063/1.4891248} {\bibfield  {journal} {\bibinfo
  {journal} {Applied Physics Letters}\ }\textbf {\bibinfo {volume} {105}},\
  \bibinfo {pages} {043502} (\bibinfo {year} {2014})}\BibitemShut {NoStop}%
\bibitem [{\citenamefont {Karlen}\ \emph {et~al.}(2019)\citenamefont {Karlen},
  \citenamefont {Haesler}, \citenamefont {Overstolz}, \citenamefont
  {Bergonzi},\ and\ \citenamefont {Lecomte}}]{karlenSealingMEMSAtomic2019}%
  \BibitemOpen
  \bibfield  {author} {\bibinfo {author} {\bibfnamefont {S.}~\bibnamefont
  {Karlen}}, \bibinfo {author} {\bibfnamefont {J.}~\bibnamefont {Haesler}},
  \bibinfo {author} {\bibfnamefont {T.}~\bibnamefont {Overstolz}}, \bibinfo
  {author} {\bibfnamefont {G.}~\bibnamefont {Bergonzi}}, \ and\ \bibinfo
  {author} {\bibfnamefont {S.}~\bibnamefont {Lecomte}},\ }\bibfield  {title}
  {\enquote {\bibinfo {title} {Sealing of {{MEMS Atomic Vapor Cells Using Cu-Cu
  Thermocompression Bonding}}},}\ }\href {\doibase 10.1109/JMEMS.2019.2949349}
  {\bibfield  {journal} {\bibinfo  {journal} {Journal of Microelectromechanical
  Systems}\ }\textbf {\bibinfo {volume} {29}},\ \bibinfo {pages} {1--5}
  (\bibinfo {year} {2019})}\BibitemShut {NoStop}%
\bibitem [{\citenamefont {Maurice}\ \emph {et~al.}(2022)\citenamefont
  {Maurice}, \citenamefont {Carl{\'e}}, \citenamefont {Keshavarzi},
  \citenamefont {Chutani}, \citenamefont {Queste}, \citenamefont
  {{Gauthier-Manuel}}, \citenamefont {Cote}, \citenamefont {Vicarini},
  \citenamefont {Abdel~Hafiz}, \citenamefont {Boudot},\ and\ \citenamefont
  {Passilly}}]{mauriceWaferlevelVaporCells2022}%
  \BibitemOpen
  \bibfield  {author} {\bibinfo {author} {\bibfnamefont {V.}~\bibnamefont
  {Maurice}}, \bibinfo {author} {\bibfnamefont {C.}~\bibnamefont {Carl{\'e}}},
  \bibinfo {author} {\bibfnamefont {S.}~\bibnamefont {Keshavarzi}}, \bibinfo
  {author} {\bibfnamefont {R.}~\bibnamefont {Chutani}}, \bibinfo {author}
  {\bibfnamefont {S.}~\bibnamefont {Queste}}, \bibinfo {author} {\bibfnamefont
  {L.}~\bibnamefont {{Gauthier-Manuel}}}, \bibinfo {author} {\bibfnamefont
  {J.-M.}\ \bibnamefont {Cote}}, \bibinfo {author} {\bibfnamefont
  {R.}~\bibnamefont {Vicarini}}, \bibinfo {author} {\bibfnamefont
  {M.}~\bibnamefont {Abdel~Hafiz}}, \bibinfo {author} {\bibfnamefont
  {R.}~\bibnamefont {Boudot}}, \ and\ \bibinfo {author} {\bibfnamefont
  {N.}~\bibnamefont {Passilly}},\ }\bibfield  {title} {\enquote {\bibinfo
  {title} {Wafer-level vapor cells filled with laser-actuated hermetic seals
  for integrated atomic devices},}\ }\href {\doibase
  10.1038/s41378-022-00468-x} {\bibfield  {journal} {\bibinfo  {journal}
  {Microsystems \& Nanoengineering}\ }\textbf {\bibinfo {volume} {8}},\
  \bibinfo {pages} {1--11} (\bibinfo {year} {2022})}\BibitemShut {NoStop}%
\bibitem [{\citenamefont {Seltzer}\ \emph {et~al.}(2008)\citenamefont
  {Seltzer}, \citenamefont {Rampulla}, \citenamefont {{Rivillon-Amy}},
  \citenamefont {Chabal}, \citenamefont {Bernasek},\ and\ \citenamefont
  {Romalis}}]{seltzer_testing_2008}%
  \BibitemOpen
  \bibfield  {author} {\bibinfo {author} {\bibfnamefont {S.~J.}\ \bibnamefont
  {Seltzer}}, \bibinfo {author} {\bibfnamefont {D.~M.}\ \bibnamefont
  {Rampulla}}, \bibinfo {author} {\bibfnamefont {S.}~\bibnamefont
  {{Rivillon-Amy}}}, \bibinfo {author} {\bibfnamefont {Y.~J.}\ \bibnamefont
  {Chabal}}, \bibinfo {author} {\bibfnamefont {S.~L.}\ \bibnamefont
  {Bernasek}}, \ and\ \bibinfo {author} {\bibfnamefont {M.~V.}\ \bibnamefont
  {Romalis}},\ }\bibfield  {title} {\enquote {\bibinfo {title} {Testing the
  effect of surface coatings on alkali atom polarization lifetimes},}\ }\href
  {\doibase 10.1063/1.2985913} {\bibfield  {journal} {\bibinfo  {journal}
  {Journal of Applied Physics}\ }\textbf {\bibinfo {volume} {104}},\ \bibinfo
  {pages} {103116} (\bibinfo {year} {2008})}\BibitemShut {NoStop}%
\bibitem [{\citenamefont {Pitz}, \citenamefont {Wertepny},\ and\ \citenamefont
  {Perram}(2009)}]{pitz_pressure_2009}%
  \BibitemOpen
  \bibfield  {author} {\bibinfo {author} {\bibfnamefont {G.~A.}\ \bibnamefont
  {Pitz}}, \bibinfo {author} {\bibfnamefont {D.~E.}\ \bibnamefont {Wertepny}},
  \ and\ \bibinfo {author} {\bibfnamefont {G.~P.}\ \bibnamefont {Perram}},\
  }\bibfield  {title} {\enquote {\bibinfo {title} {Pressure broadening and
  shift of the cesium {D}$_1$ transition by the noble gases and {N}$_2$,
  {H}$_2$, {HD}, {D}$_2$, {CH}$_4$, {C}$_2${H}$_6$, {CF}$_4$, and
  $^{3}${He}},}\ }\href {\doibase 10.1103/PhysRevA.80.062718} {\bibfield
  {journal} {\bibinfo  {journal} {Physical Review A}\ }\textbf {\bibinfo
  {volume} {80}},\ \bibinfo {pages} {062718} (\bibinfo {year}
  {2009})}\BibitemShut {NoStop}%
\end{thebibliography}%

\end{document}